\documentclass[oribibl]{llncs}
\usepackage{graphicx, amsmath, amssymb, citesort}
\usepackage{amsfonts, enumerate}

\newcommand{\A}{\mathcal{W}}

\title {Randomized Fast Design of Short DNA Words\thanks{Supported in part by NSF Grant EIA-0112934.}}
\author {Ming-Yang Kao
         \and
         Manan Sanghi
        \and
        Robert  Schweller
}

\institute{Department of Computer Science\\
Northwestern University\\ Evanston, IL 60201, USA\\
\email{\{kao,manan,schwellerr\}@cs.northwestern.edu}}

\begin{document}
\maketitle

\begin{abstract}
We consider the problem of efficiently designing sets (codes) of
equal-length DNA strings (words) that satisfy certain
combinatorial constraints. This problem has numerous motivations
including DNA computing and DNA self-assembly.  Previous work has
extended results from coding theory to obtain bounds on code size
for new biologically motivated constraints and has applied
heuristic local search and genetic algorithm techniques for code
design. This paper proposes a natural optimization formulation of
the DNA code design problem in which the goal is to design $n$
strings that satisfy a given set of constraints while minimizing
the length of the strings. For multiple sets of constraints, we
provide high-probability algorithms that run in time polynomial in
$n$ and any given constraint parameters, and output strings of
length within a constant factor of the optimal.  To the best of
our knowledge, this work is the first to consider this type of
optimization problem in the context of DNA code design.
\end{abstract}

\begin{table*}[t]
\begin{center}
\begin{tabular}{|c||c|c||c|c||}

\hline

\multicolumn{5}{|c|}{Word Length and Time Complexity for DNA Word Design}\\
\hline

\multicolumn{1}{|c||}{} & \multicolumn{1}{|c}{} &
\multicolumn{1}{c||}{} & \multicolumn{1}{|c}{} &
\multicolumn{1}{c||}{}
\\

& \multicolumn{2}{|c||}{Word Length} & \multicolumn{2}{|c||}{Time Complexity}\\

& \multicolumn{1}{|c}{\scriptsize{Lower Bound}} & \scriptsize{Upper
Bound} & \multicolumn{1}{|c}{\scriptsize{Lower Bound}} &
\scriptsize{Upper Bound}
\\ \hline

& \multicolumn{2}{|c||}{} & \multicolumn{2}{|c||}{}\\

$\mathrm{DWD}_{1,2,3,4,5,6,7}$ &
\multicolumn{2}{|c||}{$\Theta(\ell)$} &
\multicolumn{2}{|c||}{$\Theta(n\ell)$}\\

 & \multicolumn{1}{|c}{(Thm. \ref{T:lower_length})} & (Thm.
\ref{T:GC_content}) & \multicolumn{1}{|c}{(Thm.
\ref{T:lower_length})} & \multicolumn{1}{c||}{(Thm.
\ref{T:GC_content})}
\\ \hline

& \multicolumn{2}{|c||}{} & \multicolumn{2}{|c||}{}\\

$\mathrm{DWD}_{1,2,3,7,8}$ & \multicolumn{2}{|c||}{$\Theta(\ell)$} &
\multicolumn{2}{|c||}{$\Theta(n\ell)$}\\

 & \multicolumn{1}{|c}{(Thm. \ref{T:lower_length})} & (Thm.
\ref{T:consecutive}) & \multicolumn{1}{|c}{(Thm.
\ref{T:lower_length})} & \multicolumn{1}{c||}{(Thm.
\ref{T:consecutive})}
\\ \hline

& \multicolumn{2}{|c||}{} & &\\

$\mathrm{DWD}_{1,2,3,4,5,6,9}$ &
\multicolumn{2}{|c||}{$\Theta(\ell)$} & $\Omega(n\ell)$ & $O\bigg(
\min \left\{
\begin{array}{l}
   \ell^{1.5} \log^{0.5}\ell + n\ell,\\ n\ell\log\ell
\end{array}
\right\}\bigg)$\\


 & \multicolumn{1}{|c}{(Thm. \ref{T:lower_length})} & (Thm.
\ref{T:constraints8}) & (Thm. \ref{T:lower_length}) & (Thm.
\ref{T:constraints8})
\\ \hline

\end{tabular}
\end{center}
\vspace{-2ex} \caption{This table summarizes our results regarding
the efficient design of DNA words.  Here $n$ is the number of
words; $k$ denotes the maximum of the constraint parameters for
constraints 1 through 6 (see Section~\ref{S:prelims}); and
$\ell=\Theta(k+\log n)$ denotes the optimal achievable word length
for the listed word design problems (see
Theorems~\ref{T:lower_length}, \ref{T:GC_content},
\ref{T:consecutive} and \ref{T:constraints8}).} \small
\label{table:summary} \vspace{-3ex}
\end{table*}

\section{Introduction}

In this paper we study the problem of efficiently designing sets
(codes) of DNA strings (words) of near optimal length that fulfill
certain combinatorial constraints.  Many applications have emerged
in recent years that depend on the scalable design of such words.
One such problem is in DNA computing where inputs to computational
problems are encoded into DNA strands for the purpose of computing
via DNA complementary binding \cite{Adleman:1994:MCS}. Another
application involves implementing Wang tile self-assembly systems
by encoding glues of Wang tiles into strands of DNA
\cite{Winfree:1998:DSA}. DNA words can also be used to store
information at the molecular level \cite{Brenner:1997:MSP}, act as
molecular bar codes for identifying molecules in complex libraries
\cite{Brenner:1992:ECC,Brenner:1997:MSP,Shoemaker:1996:QPA}, or
implement DNA arrays \cite{BenDor:2000:UDT}.

For a set of DNA  words to be effective for the above
applications, they must fulfill a number of combinatorial
constraints.  Of particular importance is the need for specific
hybridization between a given  word and its unique Watson-Crick
complement.  That is, we need to make sure that hybridization does
not occur among a  word and the complement of a different word in
the set, or even of any  word with any other word in the set. For
this requirement Marathe et al.~\cite{Marathe:2001:CDW} have
proposed the \emph{basic Hamming constraint}, \emph{reverse
complement Hamming constraint}, and \emph{self-complementary
constraint}. We further consider the more restricting
\emph{shifting Hamming constraint} which requires a large Hamming
distance between all alignments of any pair of
words~\cite{Brenneman:2001:SDB}.

We also consider three constraints not related to Hamming distance.
The \emph{consecutive base constraint} limits the length of any run
of identical bases in any given word.  Long runs of identical bases
are considered to cause hybridization
errors~\cite{Tsaftaris:2004:DCS,Brenneman:2001:SDB}. The \emph{GC
content constraint} requires that a large percentage of the bases in
any given word are either G or C.  This constraint is meant to give
each string similar thermodynamic
properties~\cite{Tsaftaris:2004:DCS,Tulpan:2002:SLS,Tulpan:2003:HRN}.
The \emph{free energy constraint} requires that the difference in
free energy of any two words is bounded by a small constant. This
helps ensure that each word in the set has a similar melting
temperature~\cite{Brenneman:2001:SDB,Marathe:2001:CDW}.

In addition to the above constraints, it is desirable for the
length $\ell$ of each word to be as small as possible.  The
motivation for minimizing $\ell$ is evident from the fact that it
is more difficult to synthesize longer strands. Similarly, longer
DNA strands require more DNA to be used for the respective
application.

There has been much previous work in the design of DNA words
\cite{Brenneman:2001:SDB,Marathe:2001:CDW,Brenner:1997:MSP,Deaton:1996:GSR,Frutos:1997:DWD,Garzon:1997:NMD,Shoemaker:1996:QPA,Tulpan:2002:SLS,Tulpan:2003:HRN}.
In particular, Marathe et al.~\cite{Marathe:2001:CDW} have
extended results from coding theory to obtain bounds on code size
for various biologically motivated constraints.  However, most
work in this area has been based on heuristics, genetic
algorithms, and stochastic local searches that do not provide
provably good words provably fast.

In this work we provide algorithms with analytical guarantees for
combinatorial structures and time complexity. In particular, we
formulate an optimization problem that takes as input a desired
number of strings $n$ and produces $n$ length-$\ell$ strings that
satisfy a specified set of constraints, while at the same time
minimizing the length $\ell$. We restrict our solution to this
problem in two ways.  First, we require that our algorithms run in
time only polynomial in the number of strings $n$ as well as any
given constraint parameters. Second, we require that our
algorithms produce sets of words that achieve word length $\ell$
that is within a constant multiple of the optimal achievable word
length, while at the same time fulfilling the respective
constraints with high probability.  For various subsets of the
constraints we propose, we provide algorithms that do this. We
thus provide fast algorithms for the creation of sets of short
words.

\paragraph{Paper Layout:}  In Section~\ref{S:prelims}, we describe
the different biologically motivated combinatorial constraints we
use.  In Section~\ref{S:algorithms} we solve the design problem with
subsets of constraints including the Hamming constraints, the
consecutive bases constraint, and the GC content constraint.  In
Section~\ref{S:freeenergyconstraint} we extend our algorithms to
deal with the free energy constraint.

\section{Preliminaries}
\label{S:prelims}

\subsection{Notations}

Let $X=x_1x_2\ldots x_\ell$ be a  word where $x_i$ belongs to some
alphabet $\Pi$. In this paper we deal with two alphabets, namely,
the binary alphabet $\Pi_B = \{0,1\}$ and the DNA alphabet $\Pi_D$ =
\{A,C,G,T\}. The elements of an alphabet are called
\emph{characters}. We will use capital letters for words and small
letters for characters. Our goal is to design DNA words but some of
our algorithms generate binary words in intermediate steps.

The {\em reverse} of $X$, denoted by $X^R$, is the word $x_\ell
x_{\ell-1}\ldots x_1$. The {\em complement} of a character $x$ is
denoted by $x^c$. The complements for the binary alphabet are given
by $0^c = 1$, $1^c = O$, and for the DNA alphabet we have $A^c=T$,
$C^c=G$, $G^c=C$, $T^c=A$.

 The \emph{complement} of a word is
obtained by taking the complement of each of the characters in the
word, i.e., $X^C = x_1^cx_2^c\ldots x_\ell^c$.  The \emph{reverse
complement} of $X$ is the complement of $X^R$, $X^{RC} = x_{\ell}^c
x_{\ell-1}^c\ldots x_1^c$. The \emph{Hamming distance} $H(X,Y)$
between two words $X$ and $Y$ is the number of positions where $X$
differs from $Y$.

We are interested in designing a set $\A$ of $n$ words over
$\Pi_D$ each of length $\ell$ which satisfy the constraints
defined in Section \ref{SS:constraints} below.

\subsection{Constraints}\label{SS:constraints}

The constraints we consider can be classified into two categories:
\emph{non-interaction} constraints and \emph{stability}
constraints. Non-interaction constraints ensure that unwanted
hybridizations between two DNA strands are avoided,
and stability constraints ensure that the DNA strands are stable
in a solution. The first six constraints below are non-interaction
constraints while the remaining three are stability constraints.

\begin{enumerate}[${C}_1$]
    \item \hspace{-.18cm}($k_1$): \textbf{Basic Hamming Constraint} ($k_1$) = for any words $Y,X \in
    \A$, $H(Y,X)$ $\geq k_1$.

    This constraint limits
    non-specific hybridizations between the Watson-Crick
    complement of some word $Y$ with a distinct word $X$.

    \item \hspace{-.18cm}($k_2$): \textbf{Reverse Complementary Constraint} ($k_2$) = for any words $Y,X \in
    \A$, $H(Y,X^{RC}) \geq k_2.$

    This
    constraint is intended to limit hybridization between a
    word and the reverse of another word.

    \item \hspace{-.18cm}($k_3$): \textbf{Self Complementary Constraint} ($k_3$) = for any word
    $Y$, $H(Y,Y^{RC}) \geq k_3$.

    This constraint prevents a word
    from hybridizing with itself.

    \item \hspace{-.18cm}($k_4$): \textbf{Shifting Hamming Constraint} ($k_4$) = for any two words $Y,X \in
    \A$, $$H(Y[1..i],X[(\ell-i+1)..\ell]) \geq k_4 - (\ell -
    i) \mbox{ for all } i.$$ This is a stronger version of the
    Basic Hamming Constraint.

 \item \hspace{-.18cm}($k_5$): \textbf{Shifting Reverse Complementary Constraint} ($k_5$) = for any two words $Y,X \in
    \A$, $$H(Y[1..i],X[1..i]^{RC}) \geq k_5 - (\ell - i)  \mbox{ for all }
    i; \mbox{and}$$
    $$H(Y[(\ell-i+1)..\ell],X[(\ell-i+1)..\ell]^{RC}) \geq k_5 - (\ell - i)  \mbox{ for all } i.$$
    This is a stronger version of the Reverse Complementary Constraint.

 \item \hspace{-.18cm}($k_6$): \textbf{Shifting Self Complementary Constraint} ($k_6$) = for any word $Y \in
    \A$, $$H(Y[1..i],Y[1..i]^{RC}) \geq k_6 - (\ell - i)  \mbox{ for all } i; \mbox{and}$$
    $$H(Y[(\ell-i+1)..\ell],Y[(\ell-i+1)..\ell]^{RC}) \geq k_6 - (\ell - i)  \mbox{ for all } i.$$
    This is a stronger version of the Self Complementary Constraint.

    \item \hspace{-.18cm}($\gamma$): \textbf{GC Content Constraint} ($\gamma$) = $\gamma$ percentage of bases in any word
    $Y \in \A$ are either G or C.

    The GC content
    affects the thermodynamic properties of a word~\cite{Tsaftaris:2004:DCS,Tulpan:2002:SLS,Tulpan:2003:HRN}. Therefore,
    having the same ratio of GC content for all the words will
    assure similar thermodynamic characteristics.

    \item \hspace{-.18cm}($d$): \textbf{Consecutive Base Constraint} ($d$) = no word
    has more than $d$ consecutive bases for $d\geq2$.

    In some applications, consecutive
    occurrences (also known as runs) of the same base increase the
    number of annealing errors.

    \item \hspace{-.18cm}($\sigma$): \textbf{Free Energy Constraint} ($\sigma$) = for any two
    words $Y,X \in \A$, $ \mathrm{FE}(Y) -\mathrm{FE}(X) \leq \sigma$ where
    $\mathrm{FE}(W)$ denotes the free energy of a word defined in Section
   ~\ref{S:freeenergyconstraint}.

    This constraint ensures that all the words in the set have
    similar melting temperatures which allows hybridization of
    multiple DNA strands to proceed simultaneously \cite{Shoemaker:1996:QPA}.
\end{enumerate}

For each of the given constraints above we assign a shorthand
boolean function $C_i(t)$ to denote whether or not a given set of
words $\A$ fulfills constraint $C_i$ with respect to parameter $t$.
For a given integer $n$, the goal of DNA word design is to
efficiently create a set of $n$ length-$\ell$ words such that a
given subset of the above constraints are satisfied, while trying to
minimize $\ell$.  That is, for a given subset of constraints
$\{C_{\pi_1},C_{\pi_2},\ldots , C_{\pi_r} \} \subseteq
\{C_1,C_2,\ldots,C_9\}$, the corresponding DNA word design (DWD)
optimization problem is as follows.

\begin{problem}[$\mathrm{DWD}_{\pi_1,\pi_2,
\ldots,\pi_r}$]\

\noindent \textsc{Input}: Integers $n,t_1,t_2,\ldots,t_r$.

\noindent \textsc{Output}: A set $\A$ of $n$ DNA strings each of
the minimum length such that for all $1\leq i \leq r$ the
constraint $C_{\pi_i}(t_i)$ is satisfied over set $\A$.

\end{problem}

For this problem we have the following trivial lower bounds for
time complexity and the word size $\ell$ when any one of the first
six constraints is applied.

\begin{theorem}\label{T:lower_length}  Consider a set $\A$ of $n$ DNA words each of length $\ell$.
\begin{enumerate}
\item If $\A$ fulfills any one
of the constraints $C_1(k),C_2(k),C_3(k),C_4(k)$,$C_5(k)$, and
$C_6(k)$, then $\ell=\Omega(k + \log n)$.
\item The time complexity of producing a set $\A$ that fulfills any one
of the constraints $C_1(k)$, $C_2(k)$, $C_3(k)$, $C_4(k)$, $C_5(k)$,
and $C_6(k)$ is $\Omega(nk + n\log n)$.
\end{enumerate}
\end{theorem}

The goal of DNA word design is to simultaneously satisfy as many of
the above nine constraints as possible while achieving words within
a constant factor of the optimal length $\ell$ for the given set of
constraints.  In Section~\ref{S:algorithms} we show how to
accomplish this goal for various subsets of the constraints.

\section{Algorithms for DNA Word Design}
\label{S:algorithms} In this section we develop randomized
algorithms to generate sets of length-$\ell$ DNA words that satisfy
certain sets of constraints while keeping $\ell$ within a constant
of the optimal value.  In particular, we first show how simply
generating a set of $n$ words at a specific length $\ell = O(k +
\log n)$ uniformly at random is sufficient to fulfill constraints 1,
2, 3, 4, 5, and 6 simultaneously with high probability.  We then
propose three extensions to this algorithm to fulfill different
subsets of constraints within a constant factor of the optimal word
length.  The first extension yields an algorithm for fulfilling the
GC content constraint while the second yields one for the
consecutive base and GC content constraints at the cost of the
shifting constraints. Finally, we extend the basic randomized
algorithm to fulfill the free energy constraint. The first is thus
an algorithm for simultaneously fulfilling constraints 1, 2, 3, 4,
5, 6, and 7, the second simultaneously fulfills constraints 1, 2, 3,
7, and 8, and the last one fulfills constraints 1, 2, 3, 4, 5, 6 and
9.

\subsection{A Simple Randomized Algorithm}
\label{subsection:simplerandomizedsheme}

\begin{problem}[$\mathrm{DWD}_{1,2,3,4,5,6}$]\

\noindent \textsc{Input}: Integers $n$, $k_1$, $k_2$, $k_3$, $k_4$,
$k_5$, $k_6$.

\noindent \textsc{Output}: A set $\A$ of $n$ DNA strings each of the
minimum length such that the constraints $C_1(k_1)$, $C_2(k_2)$,
$C_3(k_3)$, $C_4(k_4)$, $C_5(k_5)$, $C_6(k_6)$ hold.
\end{problem}

\begin{figure}[t]
\centering
\begin{tabular}{|c|}
  \hline
\begin{minipage}{\textwidth}
\vspace{1ex} \textbf{Algorithm}
${FastDWD}_{1,2,3,4,5,6}(n,k_1,k_2,k_3,k_4,k_5,k_6)$
\vspace{-1ex}
\begin{enumerate}
\item Let $k=\max\{k_1,k_2,k_3,k_4,k_5,k_6\}$.
\item Generate a set $\A$ of
$n$ words over $\Pi_D$ of length $\ell =
9{\cdot}\max\{k,\lceil\log_4 n\rceil\}$ uniformly at random.
\item Output $\A$.
\end{enumerate}
\vspace{-1ex}
\end{minipage}\\
  \hline
\end{tabular}
\vspace{-3ex} \caption{A randomized algorithm for generating $n$ DNA
strings satisfying constraints $C_1(k_1)$, $C_2(k_2)$, $C_3(k_3)$,
$C_4(k_4)$, $C_5(k_5)$, and $C_6(k_6)$.}\label{algorithm:12345}
\end{figure}

The next theorem shows that Algorithm
$\textrm{FastDWD}_{1,2,3,4,5,6}$ $(n,k_1,k_2$, $k_3,k_4$, $k_5$,
$k_6)$ in Figure \ref{algorithm:12345} yields a polynomial-time
solution to the $DWD_{1,2,3,4,5,6}$ problem with high probability.
We omit the proof in the interest of space.

\begin{theorem}\label{T:constraints1-5}
Algorithm $\mathrm{FastDWD}_{1,2,3,4,5,6}$ produces a set $\A$ of
$n$ DNA words of optimal length $\Theta(k+\log n)$  in optimal
time $\Theta(n{\cdot}k + n{\cdot}\log n)$ satisfying constraints
$C_1(k_1)$, $C_2(k_2)$, $C_3(k_3)$, $C_4(k_4)$, $C_5(k_5)$ and
$C_6(k_6)$ with probability of failure $o(1/(n+4^k))$, where
$k=\max\{k_1$, $k_2$, $k_3$, $k_4$, $k_5$, $k_6\}$.
\end{theorem}

\begin{proof}[Sketch]
The probability that two random words violate any of the
constraints $C_1(k_1)$, $C_2(k_2)$, $C_4(k_4)$, and $C_5(k_5)$,
can be bounded using Chernoff type bounds. Similarly, we can bound
the probability of a random word violating any of the constraints
$C_3(k_3)$ and $C_6(k_6)$.

We can then apply the Boole-Bonferroni Inequaltities to yield a
bound on the probability that any pair of words in a set of $n$
random words violates constraints $C_1(k_1)$, $C_2(k_2)$,
$C_4(k_4)$, or $C_5(k_5)$; or that any single word violates
constraints $C_3(k_3)$ or $C_6(k_6)$. \qed \end{proof}

\subsection{Incorporating the GC Content Constraint into $\textrm{FastDWD}_{1,2,3,4,5,6}$}
\label{subsection:gccontent}

Now we show how to modify Algorithm $\mathrm{FastDWD
}_{1,2,3,4,5,6}$ so that it produces a set of words that also
satisfies the GC content constraint. That is, we will show how to
solve the following problem.

\begin{problem}[$\mathrm{DWD}_{1,2,3,4,5,6,7}$]\

\noindent \textsc{Input}: Integers $n$, $k_1$, $k_2$, $k_3$, $k_4$,
$k_5$,$k_6, \gamma$.

\noindent \textsc{Output}: A set $\A$ of $n$ DNA strings each of the
minimum length such that the constraints $C_1(k_1)$, $C_2(k_2)$,
$C_3(k_3)$, $C_4(k_4)$, $C_5(k_5)$, $C_6(k_6)$, $C_7(\gamma)$ hold.
\end{problem}

\begin{figure}[t]
\centering
\begin{tabular}{|c|}
  \hline
\begin{minipage}{\textwidth}
\vspace{1ex} \textbf{Algorithm}
{${FastDWD}_{1,2,3,4,5,6,7}(n,k_1,k_2,k_3,k_4,k_5,k_6,\gamma)$}\
\label{algorithm:123456} \vspace{-1ex}
\begin{enumerate}
\item Let $k=\max\{k_1,k_2,k_3,k_4,k_5,k_6\}$.
\item Generate a set $\A$
of $n$ words over the binary alphabet $\Pi_B$ of length $\ell =
10{\cdot}\max\{k,\lceil\log_2 n\rceil\}$ uniformly at random.
\item For each word $W \in \A$, for any $\lceil\gamma{\cdot}\ell\rceil$ characters in
$W$, replace $0$ by G and $1$ by C. For the remaining characters
replace $0$ by A and $1$ by T to get $W'$. Let $\A'$ be the set of
all words $W'$.
\item Output $\A'$.
\end{enumerate}
\vspace{-1ex}
\end{minipage}\\
  \hline
\end{tabular}
\vspace{-3ex} \caption{A randomized  algorithm for generating $n$
DNA strings satisfying constraints $C_1(k_1)$, $C_2(k_2)$,
$C_3(k_3)$, $C_4(k_4)$, $C_5(k_5)$, $C_6(k_6)$, and
$C_7(\gamma)$.}\label{algorithm:123456}
\end{figure}

We modify Algorithm $\textrm{FastDWD}_{1,2,3,4,5,6}$ to get
Algorithm $\textrm{FastDWD}_{1,2,3,4,5,6,7}$ shown in
Figure~\ref{algorithm:123456}. The next theorem shows that
$\textrm{FastDWD}_{1,2,3,4,5,6,7}$ yields a polynomial-time solution
to $DWD_{1,2,3,4,5,6,7}$ with high probability. We omit the proof in
the interest of space.

\begin{theorem}\label{T:GC_content}
Algorithm $\mathrm{FastDWD }_{1,2,3,4,5,6,7}$ produces a set $\A$
of $n$ DNA words of optimal length $\Theta(k+\log n)$ in optimal
time $\Theta(n{\cdot}k + n{\cdot}\log n)$ satisfying constraints
$C_1(k_1)$, $C_2(k_2)$, $C_3(k_3)$, $C_4(k_4)$, $C_5(k_5)$,
$C_6(k_6)$, and $C_7(\gamma)$ with probability of failure
$o(1/(n+2^k))$, where $k=\max\{k_1$, $k_2$, $k_3$, $k_4$, $k_5$,
$k_6\}$.
\end{theorem}

\subsection{Incorporating the Consecutive Bases Constraint into $\textrm{FastDWD}_{1,2,3,4,5,6,7}$}
\label{SS:consecutivebasesconstraints}

Now we modify Algorithm $\mathrm{FastDWD }_{1,2,3,4,5,6,7}$ so that
it produces a set that satisfies both the GC content constraint and
the consecutive base constraint at the cost of the shifting
constraints. That is, we will show how to solve the following
problem.

\begin{problem}[$\mathrm{DWD}_{1,2,3,7,8}$]\

\noindent \textsc{Input}: Integers $n$, $k_1$, $k_2$, $k_3, \gamma,
d$.

\noindent \textsc{Output}: A set $\A$ of $n$ DNA strings each of the
minimum length such that the constraints $C_1(k_1)$, $C_2(k_2)$,
$C_3(k_3)$, $C_7(\gamma)$, $C_8(d)$ hold.
\end{problem}

We use Algorithm BreakRuns shown in Figure~\ref{A:br} to break long
runs for a binary word so that it satisfies the consecutive bases
constraint with parameter $d$. Intuitively what this algorithm does
is for a given word $X$, it outputs $X'$ by inserting characters at
intervals of $d-1$ from the left and the right in a manner such that
there are no consecutive runs of length greater than $d$. We need to
add characters from both ends to ensure that $H(X,Y^{RC}) \leq
H(X',Y'^{RC})$ where $X'$ and $Y'$ are the respective outputs for
$X$ and $Y$ from BreakRuns.

\begin{figure}[t]
\centering
\begin{tabular}{|c|}
  \hline
\begin{minipage}{\textwidth}
\vspace{1ex} \textbf{Algorithm} {$BreakRuns(X,d)$}\ \vspace{-1ex}
\begin{enumerate}
\item Let $X=x_1x_2\ldots x_\ell$. For $0<i\leq \lceil\frac{\ell}{2(d-1)}\rceil - 1$, let $x'_{\ell_i}=x_{i(d-1)}^c$ and $x'_{r_i}=x_{\ell-i(d-1)}^c$.
Let $x'_{\mathrm{mid}}=x_{\lfloor \ell/2 \rfloor}^c$.
\item Output $X'=x_1\ldots x_{d-1}x'_{\ell_1}x_d\ldots x_{\lfloor\ell/2\rfloor}x'_{\mathrm{mid}}x_{\lfloor\ell/2\rfloor+1} \ldots x_{\ell-(d-1)-1}x'_{r_1}x_{\ell-(d-1)}
\ldots$ $x_\ell$.
\end{enumerate}
\ \textbf{Algorithm}
{$\textrm{FastDWD}_{1,2,3,7,8}(n,k_1,k_2,k_3,\gamma,d)$}\
\begin{enumerate}
\item Let $k=\max\{k_1,k_2,k_3\}$.
\item Generate a set $\A$
of $n$ words over the binary alphabet $\Pi_B$ of length $\ell =
10{\cdot}\max\{k,\lceil\log_2 n\rceil\}$ uniformly at random.
\item For each word $W \in \A$, let $W'=\textrm{BreakRuns}(W,d)$.  Let
$\A'$ be the set of all words $W'$.
\item For each word $W' \in \A'$, for any $\lceil\gamma{\cdot}\ell\rceil$ characters in
$W'$, replace $0$ by $G$ and $1$ by $C$. For the remaining
characters replace $0$ by $A$ and $1$ by $T$ to get $W''$.  Let
$\A''$ be the set of all words $W''$.
\item Output $\A''$.
\end{enumerate}
\vspace{-1ex}
\end{minipage}\\
  \hline
\end{tabular}
\vspace{-3ex} \caption{Algorithms for generating $n$ DNA strings
satisfying constraints $C_1(k_1)$, $C_2(k_2)$, $C_3(k_3)$,
$C_7(\gamma)$, and $C_8(d)$.}\label{A:br} \label{algorithm:12567}
\end{figure}

We modify Algorithm $\textrm{FastDWD}_{1,2,3,4,5,6,7}$ to get
Algorithm $\textrm{FastDWD}_{1,2,3,7,8}$ shown in
Figure~\ref{algorithm:12567}. The next theorem shows that
$\textrm{FastDWD}_{1,2,3,7,8}$ yields a polynomial-time solution to
$\mathrm{DWD}_{1,2,3,7,8}$ with high probability. We omit the proof
in the interest of space.

\begin{theorem}\label{T:consecutive}
Algorithm $\mathrm{FastDWD }_{1,2,3,7,8}$ produces a set $\A$ of
$n$ DNA words of optimal length $\Theta(k+\log n)$ in optimal time
$\Theta(n{\cdot}k + n{\cdot}\log n)$ satisfying constraints
$C_1(k_1)$, $C_2(k_2)$, $C_3(k_3)$, $C_7(\gamma)$, and $C_8(d)$
with probability of failure $o(1/(n+2^k))$, where
$k=\max\{k_1,k_2,k_3\}$.
\end{theorem}

\section{Incorporating the Free Energy Constraint into $\textrm{FastDWD}_{1,2,3,4,5,6}$}
\label{S:freeenergyconstraint} Now we give an alternate
modification of Algorithm $\mathrm{FastDWD }_{1,2,3,4,5,6}$ such
that the free energy constraint is satisfied.  The free-energy
$\textrm{FE}(X)$ of a DNA word $X = x_1 x_2 \ldots x_{\ell}$ is
approximated by $\textrm{FE}(X) = \textrm{correction factor} +
\sum^{\ell-1}_{i=1} \Gamma_{x_i,x_{i+1}}$, where $\Gamma_{x,y}$ is
the pairwise free energy between base $x$ and base $y$
\cite{Breslauer:1986:PDD}. For simplicity, we denote the free
energy as simply the sum $\sum^{\ell-1}_{i=1}
\Gamma_{x_i,x_{i+1}}$ with respect to a given pairwise energy
function $\Gamma$. Let $\Gamma_{\max}$ and $\Gamma_{\min}$ be the
maximum and the minimum entries in $\Gamma$ respectively. Let
$D=\Gamma_{\max}-\Gamma_{\min}$.

We now show how to satisfy the free energy constraint $C_9(\sigma)$
for a constant $\sigma=4D+\Gamma_{\max}$, while simultaneously
satisfying constraints $1, 2, 3, 4, 5,$ and $6$.  That is, we show
how to solve the following problem.


\begin{problem}[$\mathrm{DWD}_{1,2,3,4,5,6,9}$]\

\noindent \textsc{Input}: Integers $n$, $k_1$, $k_2$, $k_3$, $k_4$,
$k_5$, $k_6$.

\noindent \textsc{Output}: A set $\A$ of $n$ DNA strings each of the
minimum length such that the constraints $C_1(k_1)$, $C_2(k_2)$,
$C_3(k_3)$, $C_4(k_4)$, $C_5(k_5)$, $C_6(k_6)$,
$C_9(4D+\Gamma_{\max})$ hold.
\end{problem}


\begin{figure}
\centering
\begin{tabular}{|c|}
  \hline

\begin{minipage}{\textwidth}
\vspace{1ex} \textbf{Algorithm}
{${FastDWD}_{1,2,3,4,5,6,9}(n,k_1,k_2,k_3,k_4,k_5,k_6)$}\
\label{A:FE} \

\vspace{2ex}

Let $\hat{S}^1, \hat{S}^2, \ldots, \hat{S}^{4^m}$ be all possible
sequences of length $m=2\ell$ where $\ell$ is as defined in
Step~\ref{FE-Step-l} below such that $\textrm{FE}(\hat{S}^1) \leq
\textrm{FE}(\hat{S}^2) \leq\cdots\leq \textrm{FE}(\hat{S}^{4^m})$.
For two strings $X$ and $Y$ of respective lengths $\ell_X$ and
$\ell_Y$ where $\ell_Y$ is even, let $X\otimes Y$ be the string
$Y[1..(\ell_Y/2)]$ $X[1..\ell_X]$ $Y[(\ell_Y/2 + 1)..\ell_Y]$. Let
$\Delta=\max_i\{\textrm{FE}(\hat{S}^{i+1})-\textrm{FE}(\hat{S}^i)\}$.

\vspace{-1ex}
\begin{enumerate}
\item Let $k=\max\{k_1,k_2,k_3,k_4,k_5,k_6\}$.
\item \label{FE-Step-l} Generate a set $\A$ of
$n$ DNA words of length $\ell = 9{\cdot}\max\{k,\lceil\log_4
n\rceil\}$ uniformly at random.
\item Let $\A_{\max} = \max_{X\in\A}\{\mathrm{FE}(X)\}$ and $\A_{\min} = \min_{X\in\A}\{\mathrm{FE}(X)\}$.

\textbf{if} $\A_{\max}-\A_{\min} \leq 3D$, \textbf{then output}
$\A$.\\\textbf{else}
\item \label{stepFE} Let $\alpha=\A_{\max}+\hat{S}^1$ and
$\beta=\alpha+\Delta$. For each $S_i \in \A$, find $\hat{S}_j$
such that $\alpha \leq \mathrm{FE}(S_i)+\mathrm{FE}(\hat{S}_j)
\leq \beta$. Let $W_i'=S_i\otimes \hat{S}_j$.
\item \textbf{output} $\A'=\{W_1',\ldots,W_n'\}$.

\end{enumerate}
\vspace{-1ex}
\end{minipage}\\
  \hline
\end{tabular}
\vspace{-3ex}\caption{A randomized algorithm for generating $n$ DNA
strings satisfying constraints $C_1(k_1)$, $C_2(k_2)$, $C_3(k_3)$,
$C_4(k_4)$, $C_5(k_5)$, $C_6(k_6)$, and
$C_9(4D+\Gamma_{\max})$.}\label{algorithm:123458}
\end{figure}

\small{We modify Algorithm} $\textrm{FastDWD}_{1,2,3,4,5,6}$ to get
Algorithm $\textrm{FastDWD}_{\small{1,2,3,4,5,6,9}}$ shown in
Figure~\ref{algorithm:123458} for solving $DWD_{1,2,3,4,5,6,9}$.
The following lemmas identify the properties of symbols
$\Delta,\A,\A_{\max},\A_{\min},S_i,\hat{S}_j,\alpha,\beta$, and
$W_i'$ defined in Figure~\ref{algorithm:123458} and are used for
proving the correctness of Algorithm
$\mathrm{FastDWD}_{1,2,3,4,5,6,9}$.



\begin{lemma}\label{L:FE-max-jump} $\Delta < 2D$.
\end{lemma}

\begin{lemma}\label{L:FE1}If $\A_{\max}-\A_{\min} > 3D$, then $\A_{\max}-\A_{\min}+2D  \leq
\mathrm{FE}(\hat{S}^{4^m})-\mathrm{FE}(\hat{S}^1)$.
\end{lemma}

\begin{lemma}\label{L:FE-exists} For each $S_i \in \A$, there exists $\hat{S}_j$ such that $\alpha \leq
\mathrm{FE}(S_i)+\mathrm{FE}(\hat{S}_j) \leq \beta$.
\end{lemma}

\begin{lemma}\label{L:FE_const}For all i, $\alpha-D \leq \mathrm{FE}(W_i') \leq \beta + D + \Gamma_{\max}$.
\end{lemma}

Section \ref{subsection:computingenergyestrings} discusses the
details for Step~\ref{stepFE} of the algorithm. Finally,
Section~\ref{SS:puttingitalltogether} establishes its correctness
and time complexity.

\subsection{Computing Strings with Bounded Energies}\label{subsection:computingenergyestrings}
In Step~\ref{stepFE} of Algorithm $\mathrm{FastDWD
}_{1,2,3,4,5,6,9}$ we need to produce a set of $n$ DNA strings
$\hat{S}_1, \hat{S}_2, \ldots \hat{S}_n$, each of a given length
$L=m$, such that $A_i \leq \mathrm{FE}(\hat{S}_i) \leq B_i$ for some
$A_i$, $B_i$ such that $B_i - A_i \leq \Delta$. That is, we need to
solve the following problem.

\begin{problem}[Bounded-Energy Strand Generation]\label{P:energies}\

\noindent \textsc{Input}:
\begin{enumerate}
\item Integers $A_i$ and
$B_i$ for $i=1$ to $n$ such that
\begin{enumerate}
   \item $A_i \geq \A_{\min}$;
   \item $B_i \leq \A_{\max}$;
   \item $B_i - A_i \leq \Delta$.
\end{enumerate}
\item Length $L$.
\end{enumerate}

\noindent \textsc{Output}:  Strings $\hat{S}_1, \hat{S}_2, \ldots
\hat{S}_n$ each of length $L$ and respective energy $E_i$ such
that $A_i \leq E_i \leq B_i$.
\end{problem}

Our solution to this problem involves transforming the blunt of the
computational task into the problem of polynomial multiplication.
Consider the following polynomial.

\begin{definition}
For any integer $\ell \geq 1$, let $f_{\ell,a,b}(x)$ be the
polynomial $\sum^{\ell\cdot m}_{z=0} \zeta_z x^z$ where coefficient
$\zeta_z$ is the number of length-$\ell$ strings whose first
character is $a$, last character is $b$, and free energy is $z$.
\end{definition}

For $f_\ell(x) = \sum_{\forall a,b\in\Pi} f_{\ell,a,b}(x)$ the
coefficient of $x^i$ denotes the number of strings of length
$\ell$ and free energy $i$.  As a first step towards our solution,
we use a subroutine BUILD($L$) which computes $\Phi$, the
polynomials $f_{L,a,b}(x)$, $f_{\lfloor L/2\rfloor,a,b}(x), \ldots
, f_{1,a,b}(x)$, for all $a,b\in\Pi$ in $O(L\log L)$ time. The
efficient computation of these polynomials relies on the following
recursive property.

\begin{lemma}\label{lemma:recursive}
For any integers $\ell_1,\ell_2 \geq 1$, $$f_{\ell_1+\ell_2,a,b}(x)
= \sum_{d_1, d_2 \in \Pi} f_{\ell_1,a,d_1}(x) \cdot
f_{\ell_2,d_2,b}(x) \cdot x^{\Gamma_{d_1,d_2}}.$$
\end{lemma}

The problem of determining the number of strings of length $L$ and
free energy $E$ is considered in \cite{Marathe:2001:CDW} and a
dynamic programming based $O(L^2)$-time algorithm is provided.
However, exploiting the recursive property of
Lemma~\ref{lemma:recursive} and Fast Fourier Transforms
\cite{Cormen:2001:IA} for polynomial multiplication the subroutine
BUILD solves this problem in faster $O(L\log L)$ time and may be
of independent interest.

Our algorithm for Problem~\ref{P:energies} has two phases, the
\emph{build} phase and the \emph{extract} phase. The build phase
constructs a data structure that permits the extract phase to be
executed quickly. In the extract phase, an extraction routine is
run $n$ times to output $\hat{S}_i$ for each $i\in [1,n]$.  Since
the extraction routine is executed $n$ times and the build routine
only once, the phase that constitutes the bottleneck for our
algorithm for Problem~\ref{P:energies} depends on the values of
$n$ and $L$. We thus provide two forks for the algorithm to take,
one with a fast build routine and a modestly fast extract routine,
and the other with a slower build routine but an optimally fast
extract routine.  In particular, if $n$ is sufficiently larger
than $L$, our algorithm for Problem~\ref{P:energies} calls a
routine SlowBuild(L) which improves the runtime of Extract.
Otherwise, only a faster BUILD function is called in the first
phase, leading to a slower Extract routine. The algorithm for
Problem~\ref{P:energies} is given in
Figure~\ref{A:ConstructStrings}.

\begin{figure}[t]
\centering
\begin{tabular}{|c|}
  \hline
\begin{minipage}{\textwidth}
\vspace{1ex}
\textbf{Algorithm} {${ConstructStrings}(\{A_i\},\{B_i\},
L)$}\label{A:constructStrings}\ \vspace{-1ex}
\begin{enumerate}

\item\label{alg:cs:1} Let $\Phi \gets \textrm{Build($L$)}$.

\item\label{alg:cs:2} \textbf{if} $n \geq \sqrt{\frac{L}{\log L}}$, \textbf{then}
$\Psi \gets \mathrm{SlowBuild}(L)$, \textbf{else} $\Psi \gets
\mathrm{NULL}$.

\item\label{alg:cs:3} For each $i = 1$ to $n$, find a nonzero coefficient
$\zeta_{E_i}$ of $X^{E_i}$ in some polynomial $f_{L}^{a,b}(x) \in
\Phi$ such that $A_i \leq E_i \leq B_i$.

\item\label{alg:cs:4} For $i = 1$ to $n$, set $\hat{S}_i =$ \textrm{Extract}$(E_i,\Phi,\Psi)$.

\end{enumerate}
\vspace{-1ex}
\end{minipage}
  \\
  \hline
\end{tabular}
\vspace{-3ex} \caption{This algorithm solves the Bounded Energy
Strand Generation Problem
(Problem~\ref{P:energies}).}\label{A:ConstructStrings}
\end{figure}

Algorithm ConstructStrings makes use of three subroutines -- Build,
SlowBuild and Extract. The procedure \textrm{Build}$(L)$ computes
$\Phi$, a data structure containing for all $a,b \in \Pi$ and a
given $L$, the polynomials $f_{\ell,a,b}(x)$ for $\ell = L$,
$\lfloor \frac{L}{2}\rfloor$, $\lfloor \frac{L}{4} \rfloor$,
$\lfloor\frac{L}{8} \rfloor$, $\ldots$, $1$.  This permits
Extract$(E,\Phi,\Psi)$ to obtain a length $L$ string of energy $E$
in time $O(L\log L)$.  A call to SlowBuild$(L)$ of time complexity
$O(L^{1.5}\log^{0.5}L)$ improves the complexity of
Extract$(E,\Phi,\Psi)$ to $O(L)$ by computing $\Psi$, a data
structure containing for every non-zero term $x^i$ in
$f_{\lfloor\frac{L}{2^a} \rfloor,a,b}$ a corresponding pair of
non-zero terms $x^j$ and $x^{i-j-\Gamma_{d_1,d_2}}$ in
$f_{\lfloor\frac{L}{2^{a+1}} \rfloor,a,d_1}$ and
$f_{\lfloor\frac{L}{2^{a+1}} \rfloor,d_2,b}$ respectively. This
yields the following theorem.

\begin{theorem}\label{T:COrrectness-constructStrings}
Algorithm $\mathrm{ConstructStrings}(\{A_i\}, \{B_i\}, L)$ solves
Problem~\ref{P:energies} in time \\$O(\min\{n L \log L,
L^{1.5}\log^{0.5} L + n L\})$.
\end{theorem}

\subsection{Putting it all together for $\mathrm{DWD}_{1,2,3,4,5,6,9}$}\label{SS:puttingitalltogether}

\begin{theorem}\label{T:constraints8}
Algorithm $\mathrm{FastDWD}_{1,2,3,4,5,6,9}$ produces a set of $n$
DNA words of optimal length $\Theta(k+\log n)$ in time $O(\min\{n
\ell \log \ell, \ell^{1.5}\log^{0.5} \ell + n \ell\})$ satisfying
the constraints $C_1(k_1)$, $C_2(k_2)$, $C_3(k_3)$, $C_4(k_4)$,
$C_5(k_5)$, $C_6(k_6)$, and $C_9(4D+\Gamma_{\max})$ with
probability of failure $o(1/(n+4^k))$, where $k=\max\{k_1$, $k_2$,
$k_3$, $k_4$, $k_5$, $k_6\}$.
\end{theorem}
\begin{proof}
From Theorem~\ref{T:constraints1-5} we know that $\A$ satisfies
constraints $C_1(k_1)$, $C_2(k_2)$, $C_3(k_3)$, $C_4(k_4)$,
$C_5(k_5)$, and $C_6(k_6)$ with probability of failure
$o(1/(n+4^k))$. If $\A_{\max}-\A_{\min} \leq 3D$, then
$\mathrm{FastDWD }_{1,2,3,4,5,6,9}$ outputs $\A$ which satisfies
$C_9(3D)$ and hence also satisfies $C_9(4D+\Gamma_{\max})$.
Otherwise, it is easy to verify that since $\A$ satisfies these
six constraints, so does $\A'$. From Lemma~\ref{L:FE-exists} we
know that there always exists a string $\hat{S}_j$ as required in
Step~\ref{stepFE} of $\mathrm{FastDWD }_{1,2,3,4,5,6,9}$. Further,
Lemma~\ref{L:FE_const} shows that $\A'$ satisfies
$C_9(\Delta+2D+\Gamma_{\max})$. Therefore, $\A'$ satisfies
constraints $C_1(k_1)$, $C_2(k_2)$, $C_3(k_3)$, $C_4(k_4)$,
$C_5(k_5)$, $C_6(k_6)$, and $C_9(4D+\Gamma_{\max})$ with the
stated failure probability.

The length of any word $W'\in\A'$ is at most $3\ell$ where
$\ell=\Theta(k+\log n)$, which is optimal from
Theorem~\ref{T:lower_length}.

Generating $\A$ takes $O(n{\cdot}k + n{\cdot}\log n)$ time. The bulk
of the time complexity for the algorithm comes from Step
\ref{stepFE}, which is analyzed in Section
\ref{subsection:computingenergyestrings} to get $O(\min\{n L \log
L$, $L^{1.5}\log^{0.5} L + n L\})$ (see
Theorem~\ref{T:COrrectness-constructStrings}) where $L=O(\ell)$.
\qed \end{proof}

\section{Future Work}

A number of problems related to this work remain open.  It is
still unknown how to generate words of optimal length that
simultaneously satisfy the free energy constraint and the
consecutive bases constraint.  We also have not provided a method
for combining the consecutive bases constraint with any of the
shifting constraints.

Another open research area is the verification problem of testing
whether or not a set of words satisfy a given set of constraints.
This problem is important because our algorithms only provide a
high-probability assurance of success.  While verification can
clearly be done in polynomial time for all of our constraints, the
naive method of verification has a longer runtime than our
algorithms for constructing the sets. Finding faster, non-trivial
verification algorithms is an open problem.

A third direction for future work involves considering a
generalized form of the basic Hamming constraint.  There are
applications in which it is desirable to design sets of words such
that some distinct pairs bind with one another, while others do
not \cite{Aggarwal:2004:CGM,Tsaftaris:2004:DCS}.  In this
scenario, we can formulate a word design problem that takes as
input a matrix of pairwise requirements for Hamming distances.
Determining when such a problem is solvable and how to solve it
optimally when it is are open problems.

\bibliographystyle{siam}
\bibliography{robbie3}

\end{document}